\documentstyle[12pt,twoside,epsf]{article}
\newcommand{\fcite}[1]{$^{\mbox{\tiny\cite{#1}}}$}
\begin{document}
\begin{center}
\vspace*{0.5cm}
{\bf COMPUTER SIMULATIONS OF DEFECTS IN PEROVSKITE KNbO$_3$ CRYSTALS}
\\*[3.0ex]
R.~I.~EGLITIS$^{a,b}$, E.~A.~KOTOMIN$^{b,c}$, A.~V.~POSTNIKOV$^c$,
N.~E.~CHRISTENSEN$^d$, M.~A.~KOROTIN$^e$, and G.~BORSTEL$^c$\\
$^a$Institute of Materials Research \& Engineering,
National University of Singapore, Singapore 119260;
$^b$Institute of Solid State Physics, University of Latvia,
8 Kengaraga, Riga LV-1063, Latvia;
$^c$Universit\"at Osnabr\"uck -- Fachbereich Physik,
Osnabr{\"u}ck D-49069, Germany;
$^d$Institute of Physics and Astronomy,
University of Aarhus, Aarhus C, DK-8000, Denmark;
$^e$Institute of Metal Physics, Yekaterinburg GSP-170, Russia
\\*[3ex]
\end{center}
An {\it ab initio} LMTO approach and
 semi-empirical quantum chemical INDO method have been used for
supercell calculations of basic point defects -- $F$-type centers
and hole polarons bound to cation vacancy -- in partly covalent
perovskite KNbO$_3$. We predict the existence of both one-site and
two-site (molecular) polarons with close absorption energies
($\approx$ 1 eV). The relevant experimental data are discussed and
interpreted.
\\*[2ex]
\\*[2.0ex]

\renewcommand{\baselinestretch}{1.25}\small\normalsize

\noindent{\bf INTRODUCTION}
\\*[2ex]
\noindent Perovskite KNbO$_3$ crystals are widely used in
non-linear optics and holo\-graphy. Their properties
are influenced by point defects, primarily by vacancies.
Relatively little is known about such intrinsic point
defects in KNbO$_3$. A broad absorption band around 2.7 eV has been
observed in electron-irradiated crystals and ascribed to the
$F$-type centers (O vacancy which trapped one or two
electrons -- $F^+$ and $F$ centers,
respectively).\fcite{1}
Transient optical absorption at 1.2 eV has been associated
recently,\fcite{2} in analogy with other perovskites,
with a hole polaron (a hole bound to some defect).
The ESR study of KNbO$_3$ doped with Ti$^{4+}$
gives a proof that holes could be trapped by such negatively
charged defects.\fcite{3} For example, in BaTiO$_3$,
hole polarons were also found which are bound to Na and K
alkali ions replacing Ba and thus forming a negatively charged
site attracting a hole.\fcite{4}
Primary candidates for such defects are cation vacancies. In
irradiated MgO they are known to trap one or two holes giving rise
to the V$^-$ and V$^0$ centers\fcite{5,6}
which are nothing but {\it bound hole polaron and bipolaron},
respectively. Some preliminary theoretical study has been already
done by us on $F$ centers in KNbO$_3$\fcite{PRB-F}
and on hole polarons in this material.\fcite{pss-pol}
In the present paper, we report the results
of additional computer simulations using the same two different
approaches as e.g. in Ref.~\cite{PRB-F}.
We restrict ourselves to the cubic phase of KNbO$_3$,
with the lattice constant $a_0$=4.016~{\AA}.
\\*[2ex]

\noindent{\bf METHODS}
\\*[2ex]
For the study of the ground-state atomic and electronic structure
we used the {\it ab initio} linearized muffin-tin orbital (LMTO) method
based on the local density approximation (LDA).
For structure optimizations, the full-potential version of LMTO
by van Schilfgaarde and Methfessel has been used.\fcite{fplmto}
As an extension of the analysis previously done in Ref.~\cite{PRB-F}
for the $F$ center, we applied the tight-binding LMTO method\fcite{tblmto}
in the modification incorporating the LDA+$U$ formalism.\fcite{lda+u}
The latter allows to maintain the orbital dependency of the potential
and, to some extent, to treat Coulomb correlation effects within
localized shells beyond the LDA. We used this approach for
introducing an {\it ad hoc} upward shift of the conduction band
(mostly consisting of Nb $4d$ states)
and splitting-off of the $F$ center level from it in a different way than
it was done in Ref.~\cite{PRB-F}. Moreover, this allowed us to
analyze the symmetry of the $F$ center wavefunction.
We used $U$=8 eV and $J$=0 (for the Nb $4d$ shell)
as parameters of the LDA+$U$ method.

In parallel with the LMTO, the semi-empirical
method of the Intermediate Neglect of the Differential Overlap
(INDO) modified for ionic and partly ionic solids\fcite{8} has been used
(see Ref.~\cite{9} for details of its application to KNbO$_3$).
Differently from the LMTO,
the INDO method is based on the Hartree-Fock formalism and allows
self-consistent calculations of the excited states of defects and
thus the relevant absorption energies using the so-called
$\Delta$SCF method.
The simulation of all defects has been done within a supercell
approach, with one (neutral) O atom removed to model the $F$ center
and two atoms, O and K, removed to simulate the $F^+$ center.
For the modeling of the hole centers, a K atom has been removed, and the
actual type of the hole polaron (one-site and two-site)
was set by the symmetry of the local lattice relaxation.
In LMTO calculations, the $2\!\times\!2\!\times\!2$ supercells
containing 40 atoms were used in all cases,
i.e. with repeated point defects separated by $\sim$8~{\AA}.
Only the positions of nearest neighbors to the defect
were relaxed.
In the INDO calculations we used much larger,
$4\!\times\!4\!\times\!4$ supercells (320 atoms), and allowed
for the relaxation of more distant neighbors.
Besides decreasing the residual interaction between impurities
in adjacent supercells, this effectively takes into account
the dispersion of energy bands over the Brillouin zone of KNbO$_3$
up to a better extent than it was possible in previous defect
calculations\fcite{PRB-F,pss-pol} with smaller supercells.
\\*[2ex]

\noindent{\bf RESULTS AND DISCUSSIONS}\\*[2ex]
\underline{\bf $F$-type centers}\\
In the cubic phase all O atoms are equivalent and have the local
symmetry C$_{4v}$ (due to which the excited state of the $F$-type
centers could be split into a nondegenerate and a doubly-degenerate
levels). The optimized atomic relaxation around the $F$
center as done by the LMTO indicates the outward shift of
the Nb neighbors to the O vacancy by 3.5\% a$_0$.
The associated lattice relaxation energy is shown in Table 1.

\begin{table}[th]
\begin{center}
TABLE I ~~Absorption ($E_{\mbox{\small abs}}$) and lattice
relaxation ($E_{\mbox{\small rel}}$) energies (eV) as
calculated for the electron and hole centers by LMTO and INDO
methods.

\vspace*{0.3cm}
\begin{tabular}{
l@{\hspace*{1.8cm}}c@{\hspace*{1.8cm}}c@{\hspace*{1.1cm}}c}
\hline
 & $E_{\mbox{\small abs}}$ &\multicolumn{2}{c}{$E_{\mbox{\small rel}}$}\\
                  &    INDO    & LMTO & INDO \\
\hline
$F$-center        & 2.68; 2.93 & 0.5  & 1.35 \\
$F^+$-center      & 2.30; 2.63 & ---  & 2.33 \\
one-site polaron  & 0.9        & 0.14 & 0.4  \\
two-site polaron  & 0.95       & 0.18 & 0.53 \\ \hline
\end{tabular}
\end{center}
\vspace*{-0.5cm}
\end{table}

The optimized Nb relaxation found in the INDO
simulations was 3.9\%, i.e. very close to the {\it ab initio}
calculations. The outward relaxation of nearest K atoms and inward
displacements of O atoms are much smaller. They give $\approx$20\%
of the net relaxation energy of 1.35 eV.
The $F$ center local energy level lies $\approx$0.6
eV above the top of the valence band. Its molecular orbital
contains primarily the contribution from the atomic orbitals of
the two nearest Nb atoms. Only $\approx$0.6 $e$
resides at the orbitals centered at the vacancy site; hence the
electron localization at the defect is much smaller
than is known for $F$ centers in ionic oxides
(see Ref.~\cite{10} for comparison).
The symmetry analysis of the ground-state wave function associated
with the $F$ center, done by the TB-LMTO method with the use of
the LDA+$U$ formalism and by INDO, revealed the same result,
namely that the major contribution comes from the $e_g$ states
centered at Nb neighbors (more specifically, it is essentially
the $3z^2\!-\!r^2$ component, with $z$ in the direction towards
the $F$ center). The partial densities of states from the LMTO
calculation are shown in Fig. 1.

\vspace*{6.6cm}
\setlength{\unitlength}{1.0cm}
\begin{picture}(0,0)
\put(-1.7, -4.6){
 \epsfxsize 14.0cm
 \epsffile{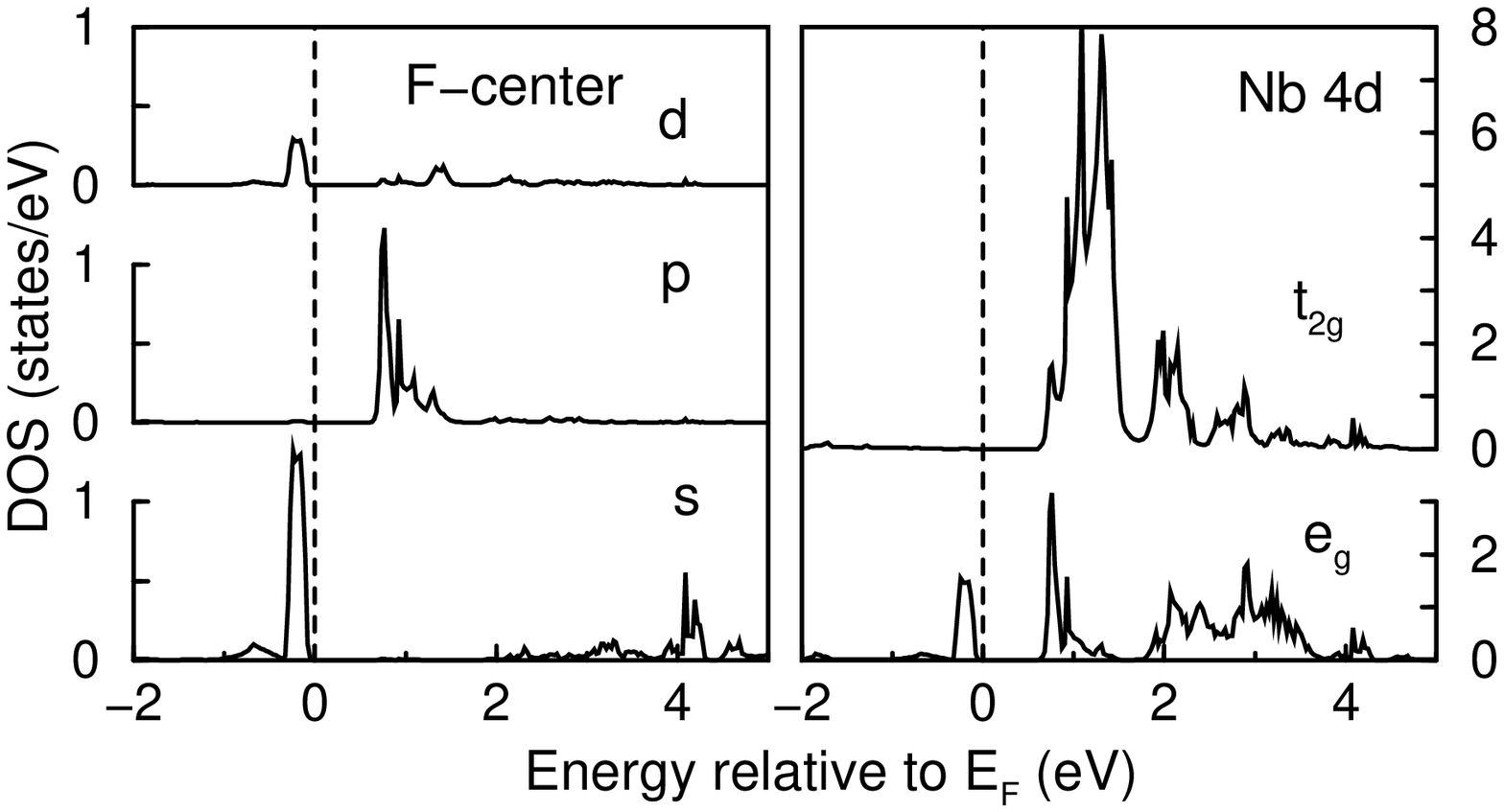}
}
\end{picture}
\vspace*{0.6cm}
\begin{center}
FIGURE 1 ~~Local density of states of the $F$-center (left panel)
and of the Nb atom nearest to it as calculated by the LMTO method.
\end{center}

For the $F^+$
center the relaxation energy of 2.23 eV and the Nb displacements
of 5.1\% are larger than those for the $F$ center due to a
stronger Coulomb repulsion between unscreened O vacancy and Nb
atoms: a share of the electron density inside the O vacancy
decreases to 0.3 $e$.

The optical absorption energies calculated by means of the
$\Delta$SCF method for the $F^+$ and $F$ centers are given in
Table 1. The two absorption bands predicted for the former center
are shifted to the low-energy side, which is in agreement with
similar defects in ionic oxides.\fcite{10} However, {\it both} defects are
predicted to have one of the bands around 2.6--2.7 eV, in
agreement with the experimental observation.\fcite{1} \\

\noindent\underline{\bf Hole Polarons}\\
Both {\it ab initio} and semi--empirical calculations agree that
there are $two$ energetically favorable atomic configurations
in which a hole is well localized: one-site and two-site
(molecular) polarons. In the former case, a single O$^-$ ion is
displaced towards the K vacancy by 1.5 \% (LMTO) or 3\% (INDO).
The INDO calculations show that simultaneously,
11 other nearest oxygens surrounding the vacancy are slightly
displaced outwards the vacancy. In the two-site
(molecular) configuration, a hole is shared by the two O atoms
which approach each other -- by 0.5\% (LMTO) or 3.5\% (INDO) -- and
both shift towards a vacancy -- by 1.1\% (LMTO) or 2.5\% (INDO).
The lattice relaxation energies (which could be associated with
the experimentally measurable hole thermal ionization energies)
are presented in Table 1. In both methods the two-site
configuration of a polaron is lower in energy.

In spite of general observation of a considerable degree of
covalency in KNbO$_3$ (see, e.g., Ref.~\cite{9} for a discussion)
and contrary to a delocalized character of the $F$ center state,
the one-site polaron state remains well localized at the
displaced O atom, with only a small contribution
from atomic orbitals of other O ions but not K or Nb ions.
In agreement with Schirmer's theory for the small-radius
polarons in ionic solids,\fcite{schirmer}
the optical absorption corresponds to a hole transfer to the state
delocalized over nearest oxygens.
The calculated absorption energies for
one-site and two-site  polarons  are close (Table 1)
and twice smaller than the experimental value
for a hole polaron trapped near Ti.\fcite{3}
This shows that the optical absorption energy of small bound polarons
could be strongly dependent on the defect involved.
\\*[2ex]

\noindent{\bf CONCLUSIONS} \\*[2ex]
\indent
(i) The two different methods used for defect calculations reveal
a qualitative agreement, despite the fact that the INDO
(as is generally typical for the Hartree-Fock-based schemes)
systematically gives larger atomic displacements and relaxation
energies. Both two--electron $F$-center calculations demonstrate a
strong electron delocalization from the O vacancy over the two
nearest Nb atoms; very likely due to a considerable covalency of
the chemical bonding in KNbO$_3$, and predict the e$_g$ symmetry of
the wave function which could be checked experimentally. Both
methods agree also that both one--site and two--site hole polarons
bound to the cation vacancy are energetically favorable, with a
preference to the latter.

(ii) The INDO calculations of the optical absorption energies strongly
support the interpretation of the experimentally observed band at 2.7
eV as due to the $F$-type centers. To distinguish between the $F$
and $F^+$ centers, the ESR measurements should be used.

(iii) The calculated hole polaron absorption ($\approx$1 eV) is close to
the observed short-lived absorption band energy\fcite{2} and thus could
arise due to a hole polaron bound at cation vacancy. Further
detailed study is needed to clarify whether such hole polarons are
responsible for the effect of the
blue-light-induced-infrared-absorption (BLIIRA) reducing the
second-harmonic generation efficiency in KNbO$_3$.\fcite{2}
\\

\noindent{\bf Acknowledgments}\\
This study was partly supported
by the DFG (a grant to E.~K.; the participation of A.~P.
and G.~B.~in the SFB 225), Volkswagen Foundation
(grant to R.~E.), and the Latvian National Program on New
Materials for Micro- and Optoelectronics (E.~K.).
M.~K. appreciates the hospitality of the University of Osnabr\"uck
during his stay there. Authors are
greatly indebted to Prof. M.~R.~Philpott for valuable discussions.

\newcommand{\myskip}[0]{\vspace*{-0.33cm}}

\end{document}